\documentclass[aps,pre,showpacs,twocolumn,groupedaddress]{revtex4}
\newcommand{\BE}{\begin{equation}}
\newcommand{\EE}{\end{equation}}
\newcommand{\BEA}{\begin{eqnarray}}
\newcommand{\EEA}{\end{eqnarray}}
\newcommand{\BA}{\begin{array}}
\newcommand{\EA}{\end{array}}
\usepackage{graphicx}
\begin{document}

\title{
Collective motions in globally coupled tent maps with stochastic updating
}

\author
{Satoru Morita}
\email{morita@sys.eng.shizuoka.ac.jp}
\affiliation{
Department of Systems Engineering,
Shizuoka University, 3-5-1 Jouhoku, Hamamatsu 432-8561, Japan 
}
\author{Tsuyoshi Chawanya}
\affiliation{
Department of Mathematics, Graduate School of Science,
Osaka University, Toyonaka 560-0043, Japan}

\begin{abstract}
We study a generalization of globally coupled maps,
where the elements are updated with probability $p$.
When $p$ is below a threshold $p_c$, the collective motion vanishes
and the system is the stationary state in the large size limit. 
We present the linear stability analysis.
\end{abstract}

%\keywords{coupled maps, collective motion, asynchronous updating}

\date{\today}
\maketitle

\section{Introduction}

The globally coupled maps (GCM) are introduced as a simple
model capturing the essential features of 
nonlinear dynamical systems with many degrees of freedom \cite{kan0}.
One of the most interesting phenomena seen in such systems is
the emergence of the collective motion \cite{kan1,chate,kuramo}.
The collective motion is  characterized by a time dependence
of the macroscopic variable in the large size 
limit \cite{pik,kan2,just,ersh,mo1,mo2,nak,cha,shib1,shib2,per0,per1}.
In this paper, we consider a variation of GCM to include
asynchronous updating.

The general form of GCM is written in the following way
\BE
x_{t+1}(i)=f(x_{t}(i))+{K\over N}\sum_{i'=1}^{N}g(x_{t}(i')) \ ,
\label{model0}
\EE
where $t$ represents discrete time steps, $i$ specifies each element,
$K$ gives the coupling strength, and $N$ is the system size.
All elements are updated synchronously in the deterministic
way through the mean field
\BE
h_t \equiv {1\over N}\sum_{i=1}^{N}g(x_{t}(i)) \ .
\label{mean-field1}
\EE
Since the collective motions 
have been studied analytically in globally coupled tent maps 
\cite{pik,kan2,just,ersh,mo1,mo2,nak,cha,shib1},
we specifically consider tent maps as follows:
\BE
\BA{lcl}
f(x) & \equiv & \displaystyle a \left (\frac{1}{2}-|x| \right ) \ ,  \\
g(x) & \equiv & \displaystyle f(x)-\overline f \ .
\EA
\label{map}
\EE
Here $\overline f$ is a constant determined from the average 
of $f(x)$ over the the natural invariant measure of the map 
$x \mapsto f(x)$, i.e.,
\BE
\overline f=\int f(x) \rho_*(x)dx \ ,
\EE
where $\rho_*(x)$ represents the the natural invariant density.
By the above choice of $g(x)$, 
$h_t=0$ is a stationary solution for the large size limit
($N\rightarrow \infty$) of (\ref{model0}),
and the corresponding stationary distribution is proportional to 
$\rho_*(x)$.
Equation (\ref{map}) looks a little different from
well-known form of GCM system
\BE
x_{t+1}(i)=(1-\epsilon)f(x_t(i))+
\frac{\epsilon}{N}\sum_i f(x_t(i)) \ ,
\label{model2}
\EE
which is obtained as a mean-field approximation 
for the coupled map lattice with diffusion coupling.
In the case of the tent map system, however,
the diffusion form (\ref{model2}) is scaled
into (\ref{model0}) \cite{cha}.

The collective motions in GCM (\ref{model0})
with (\ref{map}) are classified into two types
according to the gradient $a$ of the tent map.
First, in the case of $a<1$,
the synchronized chaos is stable.
In this case, the map $x\mapsto f(x)$ has the stable fixed point
\BE
x_*=\frac{a}{2a+2} \ .
\label{fixed_x}
\EE
Thus, $\rho_*(x)=\delta(x-x_*)$, i.e.,
$g(x)=f(x)-f(x_*)$.
Since the gradient of $f(x)$ is smaller than 1,
the difference between any pair of elements diminishes.
Thus all the elements behave identically after 
some initial transient.
Here we concentrate on the long-term behavior and assume 
the system is one-cluster state.
Then temporal evolution for $x_t$ and $h_t$ is obtained as follows 
\BE
\BA{ccc}
x_{t+1} & = & \displaystyle 
\left[\frac{1}{2}+\frac{aK}{2+2a}-(1+K)|x_t|\right] \\ 
h_{t+1} & = & a\left[x_*-\left|x_*+(1+K)h_t\right|\right] \ .\\
\label{dyn_mean-field}
\EA
\EE
When $K$ is so large that $a(1+K)>1$, 
the fixed point $h_t=0$ is unstable.
Then, the motion of the mean field is 
one-dimensional chaos, which obeys (\ref{dyn_mean-field}).

Second, in the case of $a>1$, 
all elements are fully desynchronized 
and behave as if they are mutually independent.
Nevertheless, the fluctuation of the mean field dose not vanish
in the large size limit \cite{kan1}.
Thus, the system has a nontrivial collective motion \cite{pik}.

From a realistic viewpoint, however, the synchronous updating is not
always plausible as models of real 
systems \cite{per2,rolf,abram,sinha,blok}.
In some cases, for example, neural networks, 
an independent choice of the times at which the state of a 
given element is updated should provide a better approximation.
Abramson and Zanette have numerically found that,
for globally coupled logistic map with completely stochastic updating,
the fluctuation of the mean field can vanish
in the large size limit \cite{abram}.
In this paper, we study the stochastic updating model as follows
\BE
x_{t+1}(i)=\left\{
\BA{ll}
\displaystyle f(x_{t}(i))+K h_t
& \mbox{with probability} \ p \\
\displaystyle x_{t}(i) & \mbox{with probability} \  (1-p)
\label{model1}
\EA
\right. \ .
\EE
At each time step, update the elements with probability $p$
satisfying $0<p\le1$.
When the updating rate $p$ is equal to 1, the model (\ref{model1}) 
becomes the synchronous updating model (\ref{model0}).
On the other hand, when $p$ decreases to 0,
the model (\ref{model1}) approaches the completely asynchronous
updating model.\footnote{For $p=0$, all elements are never updated.
Therefore, when we consider the limit of  $p\rightarrow +0$, 
the time $t$ must be rescaled by $p$ to keep the motions.}
Thus, the value of $(1-p)$ represents the strength of the asynchronousnism.
The purpose of this paper is to investigate how 
the collective behavior changes
when the updating probability $p$ varies, mainly by 
the linear stability analysis of the stationary state
in the large size limit.

\section{Numerical Results}

\begin{figure}
\includegraphics[scale=0.5]{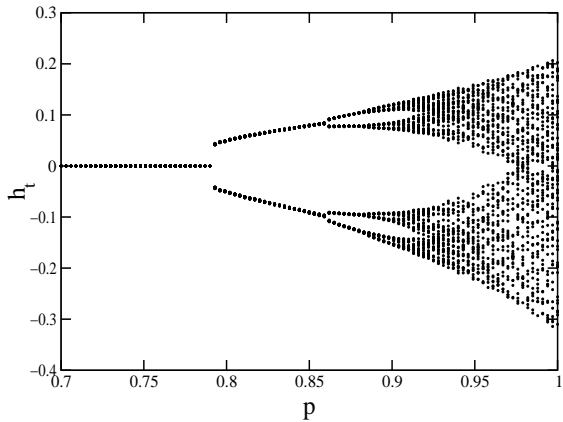}
\caption{
Bifurcation diagram for evolution of the mean field
as function of the updating rate $p$
in the case of $a=0.9, K=0.7, N=10^5$.
For given values of $p$, the value of the mean field is 
plotted for 64 successive steps 
after $10^4$ steps as transient. 
}
\label{f1}
\end{figure}

In this section, we present the numerical results for 
the model (\ref{model1}).
First, we examine the case of $a<1$.
When $p=1$, the synchronized chaos is observed.
On the other hand, when $p<1$,
some elements are updated and the others are fixed
at each time step.
Hence, even when a pair of elements have the same
value at a moment, 
they can have different values at the next time.
As a result, the synchronized chaos is broken.
When $p$ is near 1, the synchronized state is blurred slightly.
As $p$ decreases, a sequence of bifurcations are seen
(in Fig.~\ref{f1}).
It resembles the period doubling cascade.
However, there is a finite jump at this bifurcation
in contrast of the usual period doubling bifurcation.
This discontinuity is due to the fact that the map $f(x)$
is piecewise linear.
When $p$ is smaller than a threshold value $p_c$,
all elements fall into the fixed point (\ref{fixed_x})
and the mean field $h_t$ becomes 0.
Thus, the collective motion vanishes below the threshold $p_c$.

\begin{figure}
\includegraphics[scale=0.5]{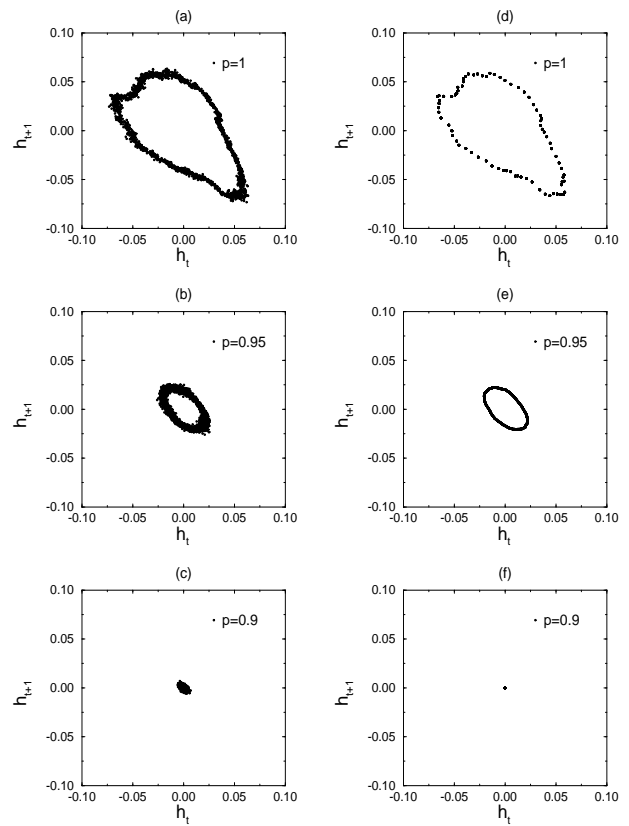}
\caption{Examples of the return maps of the mean field are shown
($a=1.625$, $K=0.4$).
While (a), (b), and (c) are obtained by the direct numerical calculation
of (\protect{\ref{model1}}) with the size of $N=10^5$,
(d), (e), and (f) are obtained by the numerical calculation of
(\protect{\ref{fp_op2}}) with $2^{18}$ grids.
(a) and (d) are for $p=1$,
(b) and (e) are for $p=0.95$,
(c) and (f) are for $p=0.9$.
We plot 2000 points after $10^4$ steps as transient.}
\label{f2}
\end{figure}

Second, we examine the case of $a>1$,
where the nontrivial collective motion is seen for $p=1$.
Figs.~\ref{f2}(a), \ref{f2}(b), and \ref{f2}(c) show
the motions of the mean field for some values of $p$,
with $a=1.625$, $K=0.4$, and $N=10^5$.
The amplitude of motion of the mean field decreases as $p$ decreases.

It should be noted that the dynamics of each element is not
deterministic due to the updating rule. 
Thus, the mean field value is blurred
if the system size is finite. 
Even for $p=1$, the finite size effect  
works on the motion of the mean field as an internal noise \cite{pik}.
For that reason, it is useful to consider the large size limit.
In the large size limit, the ensemble of the elements
is characterized by its distribution.
In the synchronous updating case,
the evolution of the distribution function $\rho_{t}(x)$ obeys
the nonlinear Frobenius-Perron equation
\BE
\BA{lcl}
\rho_{t+1}(x) & = & {\cal F}[\rho_{t}(x);h_t] \\
& \equiv & \displaystyle \frac{\rho_{t}(y_{+})+\rho_{t}(y_{-})}{a} \ ,
\label{fp_op1}
\EA
\EE
where ${\cal F}$ represents the Frobenius-Perron operator,
and $y_+$ and $y_-$ are the two preimage of $x$, i.e.,
$x=f(y_{\pm})+K h_t$. 
Here the mean field is determined in the integral form:
\BE
h_t=\int g(x)\rho_t(x)dx \ .
\label{mean-field2}
\EE
In the stochastic updating case ($p<1$),
the evolution of $\rho_t(x)$ is described as 
\BE
\rho_{t+1}(x)= p \ {\cal F}[\rho_{t}(x);h_t] \ + \ (1-p) \ \rho_{t}(x) \ .
\label{fp_op2}
\EE
Despite the stochastic updating,
the distribution function evolves in the deterministic way.

In order to calculate (\ref{fp_op2}) numerically,
we approximate the distribution function $\rho_t(x)$ 
by dividing the relevant interval of $x$ into $m$
small intervals.
The evolution of the distribution
is described by the $m\times m$ transfer matrix 
which depends on time through the mean field $h_t$.
Here we construct the transfer matrix by applying
the method by Binder and Campos \cite{binder}.
Figs.~\ref{f2}(d), \ref{f2}(e), and \ref{f2}(f) show
the the motion of the mean field 
calculated by this method with the parameter values corresponding to
Figs.~\ref{f2}(a), \ref{f2}(b), and \ref{f2}(c), respectively.
The direct calculations of (\ref{model1}) compare successfully with
the results of (\ref{fp_op2}),
except for the fluctuation due to the finite size effect.
As is seen from Fig.~\ref{f2}(f),
when $p$ is small, the collective motion vanishes like the case of $a<1$.
It should be noted that, in the stationary state, 
all elements are still scattered and behave chaotically 
in contrast to the case of $a<1$.
The fluctuation of the mean field resides for finite size systems
(Fig.~\ref{f2}(c)).

\section{Linear Stability Analysis of the Stationary States}

The result of numerical simulation indicates that
there exists a threshold value for updating rate $p_c$.
It is observed that 
the collective motion vanishes, and the stationary state 
(with distribution $\rho_t(x)=\rho_*(x)$) 
is realized for $p$ smaller than $p_c$.  
In this section, we present the linear stability analysis of
the stationary state to estimate the value of $p_c$.

\begin{figure}
\includegraphics[scale=0.5]{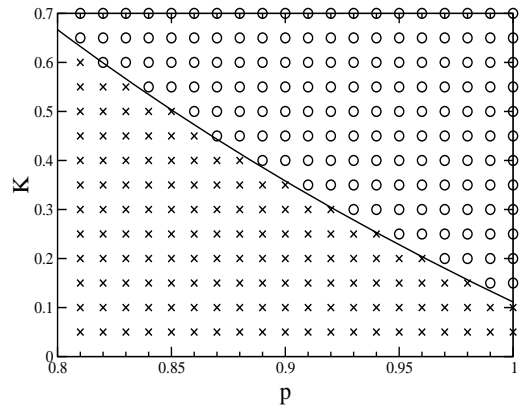}
\caption{
Phase diagram of the collective motion for $a=0.9$.
The numerical results are obtained for $N=10^4$.
Circles and crosses represent the cases of 
$\langle h^2 \rangle ^{1/2}>10^{-6}$
and $\langle h^2 \rangle ^{1/2}<10^{-6}$, respectively.
Here the angle bracket denotes the average value 
of the enclosed quantity over $2000$ time steps.
The solid line represents the theoretical prediction
(\protect{\ref{pc}}).
}
\label{f3}
\end{figure}

First, we consider the case of $a<1$.
This case is simpler, because 
all elements have the identical fixed value $x_*(>0)$ in the stationary state.
Considering a small perturbation from $x_*$,
we assume that every element has a positive value.
Since $\rho_t(x)$ for $x<0$ is 0, the evolution of $\rho_t(x)$
is rewritten as 
\BE
\rho_{t+1}(x) = \frac{p}{a} \ \rho_t
\left (\frac{a-2x+2K h_t}{2a} \right )
+ (1-p) \ \rho_t(x) \ .
\label{linear}
\EE
From (\ref{mean-field2}) and (\ref{linear}), 
we obtain the dynamics of the mean field obeys
\BE
h_{t+1}=(-a p-ap K+1-p) \ h_t \ .
\EE
Thus, the stationary state is stable when $|-ap-ap K+1-p|<1$. 
Consequently, the threshold value $p_c$ is estimated as
\BE
p_c=\frac{2}{a(1+K)+1} \ .
\label{pc}
\EE
In the case of Fig.~\ref{f2}, $p_c=0.790\cdots$.
The theoretical prediction agrees well with the numerical simulation 
(see also Fig.~\ref{f3}).

Second, we consider the case of $a>1$.
The stationary distribution function $\rho_*(x)$
is expanded into series of step functions as follows \cite{mo1,cha}
\BE
\BA{lcl}
\rho_*(x) & = & \displaystyle \sum_{k=1} C_k\theta(x-X_k) \\
X_k & \equiv & \displaystyle f^k(0) \\
C_k & \equiv & \displaystyle C_1\{(f^{k-1})'(f(0))\}^{-1} \ ,
\label{inv_meas}
\EA
\EE
where $\theta(x)$ is a step function: 1 for $x\ge 0$ and 0 for $x<0$.
Thus, $X_k$ and $C_k$ represent the locations and the heights
of the steps in $\rho_*(x)$, respectively.
We choose $C_1$ to satisfy the normalization condition
\BE
- \sum_{k=1} C_k X_k = \int dx \rho_*(x)=1 \ . 
\label{kikaku}
\EE
If there exists such $k_p$ that satisfies
$f^{k_p}(0)=0$ and $f^{k}(0)\neq0$ for $\forall k<k_p$,
the sum over $k$ is taken from 1 to $k_p$.
Otherwise, the sum is taken from 1 to $\infty$. 

Before treating the case of stochastic updating, 
let us analyze how the stationary state is affected by
adding external force with infinitesimal amplitude $\delta$
for $p=1$ and $K=0$. 
Here we assume that the external force 
changes $x_i$ for every element by the given amount.
Thus, when the force $\delta_0$ is applied at $t=0$,
the distribution function at $t=1$ is expressed as
\BE
\rho_1(x;\delta_0)=\rho_*(x-\delta_0) \ .
\label{int_cond}
\EE
In the limit of $\delta_0 \rightarrow 0$,
the response of the mean field after $\tau$ steps is written as
\BE
h_{\tau}=L_{\tau} \delta_0
\EE
where $L_{\tau}$ is the linear coefficient for the 
response with delay of $\tau$ steps \cite{cha}.
From (\ref{inv_meas}), we calculate $L_{\tau}$ as follows
\BE
L_{\tau} =  -\sum_{k=1}^{\infty} C_k X_{k+\tau} \ .
\label{prob_fn2}
\EE
When the temporal series of the external force is given
as $\{\delta_t\}$, the mean field $h_t$ is obtained as 
\BE
h_t = \sum_{\tau=1}^{\infty} L_{\tau} \delta_{t-\tau} \ ,
\label{hLc}
\EE
within the linear approximation.
Introducing a cut-off $d$, 
the state at the time $t$ can be described approximately by 
$d$-dimensional vector 
\BE
\mbox{\boldmath$v$}_t
\equiv (\delta_{t-1},\delta_{t-2},\delta_{t-3},\dots\delta_{t-d}) \ .
\EE
In order to obtain the stability condition accurately,
we must take the limit of $d\rightarrow \infty$.
When the components of $\mbox{\boldmath $v$}_t$
is denoted as $v_t^i$, we obtain 
\BE
h_t = \sum_{i=1}^{d} L_{i} \  v_t^i \ .
\label{hLc2}
\EE
The next step is to consider the case of $p=1$ and $K\neq0$.
In this case, the mean field coupling yields the feedback force.
The influence of the feedback force is described as
\BE
v_{t+1}^{1}= K h_{t} \ .
\label{cKh}
\EE
Taking (\ref{hLc2}) into account,
the evolution of $v_t^i$ is described as 
\BE
v_{t+1}^i = \sum_{j=1}^{d}J_{ij} \  v_t^j \ , 
\label{jacob_mat}
\EE
where the matrix $J_{ij}$ is given by 
\BE
J_{ij}
=
\left(
\BA{cccccc}
K L_1 & K L_2 & K L_3 & \cdots 
& K L_{\tau_c-1} & K L_{\tau_c} \\
1 & 0 & 0 & \cdots & 0 & 0 \\
0 & 1 & 0 & \cdots & 0 & 0 \\
\vdots & \vdots & \vdots & \ddots & \vdots & \vdots \\
0 & 0 & 0 & \cdots & 0 & 0 \\
0 & 0 & 0 & \cdots & 1 & 0
\EA
\right) \ . 
\label{J2}
\EE
The characteristic equation of 
the matrix (\ref{J2}) is given as 
\BE
\lambda^{d}=K\sum_{i=1}^{d}L_{i}\lambda^{d-i} \ ,
\label{eigen_eq1}
\EE
which coincides with the results of Refs. \cite{cha,keller}.
The roots of (\ref{eigen_eq1}) are denoted as
$\mu_i$.
In the case of the synchronous updating ($p=1$),
if all $\mu_{i}$ lie within the unit circle 
in the complex plane, the stationary state is stable \cite{keller}.

\begin{figure}
\includegraphics[scale=0.4]{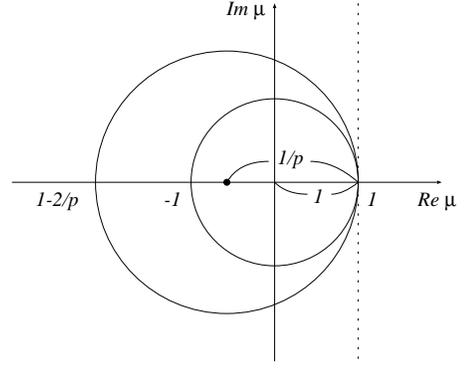}
\caption{
The stability condition.
If all solutions $\mu_i$ of (\protect{\ref{eigen_eq1}})
lie within the large (unit) circle,
the stationary state is stable for $p<1$ ($p=1$).
Thus, all $\mu_i$ lie on the left of the dotted line,
there exists threshold value $p_c$.
}
\label{f4}
\end{figure}

Let us now return to the asynchronous updating case ($p<1$).
Here we define the vector $\mbox{\boldmath $v$}_t$ to satisfy
(\ref{hLc2}).
Thus, $v_t^i$ represents the contribution from the 
past perturbations through $i$ times of updating.
Taking into account that 
the elements are updated with probability $p$
and fixed with probability $1-p$, we obtain
\BE
v_{t+1}^i = \sum_{j=1}^{d} [p J_{ij} + (1-p)\delta_{ij}]\  v_t^j \ .
\label{J4}
\EE
The characteristic equation for (\ref{J4}) is given as 
\BE
\left(\frac{\lambda-1+p}{p}\right)^{d}= K\sum_{i=1}^{d}
L_{i}\left(\frac{\lambda-1+p}{p}\right)^{d-i}
\label{eigen_eq2} .
\EE
Comparing (\ref{eigen_eq2}) with (\ref{eigen_eq1}),
$\lambda$ in (\ref{eigen_eq1}) is replaced with 
$(\lambda-1+p)/p$ in (\ref{eigen_eq2}),
and thus the stability condition for (\ref{J4}) becomes
\BE
|1-p+p\mu_i|<1 \ \mbox{(for all $i$)} \ .
\label{joken}
\EE
This condition means all $\mu_{i}$
lie within the circle with center $(1-1/p ,0)$ and radius $1/p$
in the complex plane (Fig.~\ref{f4}).
In the limit $p\rightarrow 0$, the condition (\ref{joken})
becomes $Re (\mu_i)<1$.
Consequently, if all $\mu_{i}$ satisfy $Re (\mu_i)<1$,
the system has the threshold $p_c$.

\begin{figure}
\includegraphics[scale=0.5]{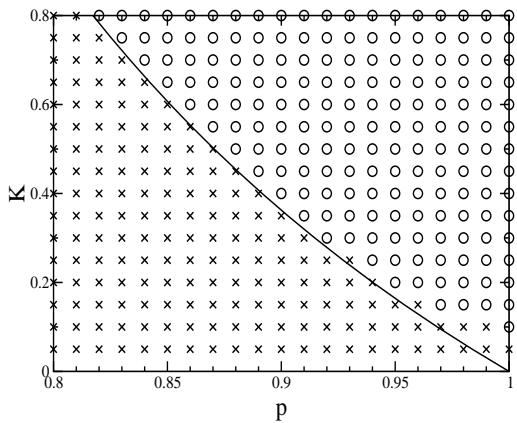}
\caption{
Phase diagram of the collective motion for 
$a=(1+ \protect{\sqrt{5}})/2$.
The numerical results are obtained by calculation
of (\protect{\ref{fp_op2}}) with $2^{18}$ grids.
Circles and crosses represent the cases of 
$\langle h^2 \rangle ^{1/2}>10^{-12}$
and $\langle h^2 \rangle ^{1/2}<10^{-12}$, respectively.
The solid line represents the theoretical prediction
(\protect{\ref{theory}}).
}
\label{f5}
\end{figure}

For example, we investigate a simple case $a=(1+\sqrt{5}/2)$,
where $f^{3}(0)=0$.
Thus, $k_p$ is 3 and $X_k$ is periodic. 
In this case, the characteristic equation (\ref{eigen_eq1})
can be solved analytically in the limit of $d\rightarrow \infty$.
From (\ref{inv_meas}) and (\ref{prob_fn2}),
the linear coefficient $\{L_{t}\}$ is given as
\BE
\BA{cll}
L_{3n+1} & = L_1 = \displaystyle - \frac{5+\sqrt{5}}{10} \ & , \\
L_{3n+2} & = L_2 = \displaystyle - \frac{5-\sqrt{5}}{10} \ & , \\
L_{3n+3} & = L_3 = 1 & . \\
\EA
\EE
Let us assume that $\mu$ satisfies condition $|\mu|>1$.
Then we rewrite the characteristic equation (\ref{eigen_eq1}) in the 
limit $d\rightarrow \infty$ as follows:
\BE
\BA{rl}
1  & = \displaystyle  K\sum_{i=1}^{\infty} L_{i} \mu^{-i} \\
& = \displaystyle K \frac{L_1 \mu^{-1} + L_2 \mu^{-2} + L_3 \mu^{-3}}
{1-\mu^{-3}} \ .
\EA
\label{CRL}
\EE
From this, $\mu$ is given as 
\BE
\mu = \  \frac{KL_1-1 \pm \sqrt{(K L_1 -1)^2 -4K-4}}{2} .
\label{kai}
\EE
When $(45-21\sqrt{5})/2<K<(5+3\sqrt{5})/2$,
the part in the square root in (\ref{kai}) is negative
and thus the solution (\ref{kai}) is complex conjugate pair. 
Then, we obtain $|\mu|=\sqrt{K+1}$.
Thus, $|\mu|>1$ holds for $0<K<(5+3\sqrt{5})/2$.
In this case, the threshold $p_c$ is given by 
\BE
p_c=\frac{K L_1-3}{K L_1-3-K}
\label{theory}
\EE
Figure \ref{f5} shows 
the correspondence between the above estimation for $p_c$
and the numerical results.
It indicates the good agreement, except for the lower right corner.
The estimation implies $p_c$ tends to 1
in the limit of $K\rightarrow 0$,
and we think that the gap in 
the corner appeared because the amplitude of the collective motion
for small $K$ is very small (estimated at $\exp(-C/K)$
for the case with $p=1$ \cite{nak,cha}). 

For such special values of $a$, where
$\{X_k\}$ falls on a periodic orbit,
we can solve the characteristic equation
by the above method.
In this case, 
the characteristic equation has no solution $\mu$ which
satisfies $Re(\mu)>1$ for adequately weak coupling. 
Thus there exists the threshold.
For the general values of $a$, however, 
it is difficult to solve the characteristic equation 
and we have not obtained the stability
condition explicitly at present.

\section{Summary and Remarks}

This study have explored the globally coupled tent maps with 
stochastic updating.
We introduced the updating rate $p$
and examined how the collective behavior changes
as $p$ varies.
In the case of $a<1$, 
the collective motion has a sequence of bifurcations,
which is similar to the period doubling cascade.
In the case of $a>1$,
the amplitude of the collective motion
decreases as $p$ decreases,
For the both cases, we observed the threshold $p_c$, 
below which the collective motion vanishes.
We estimated successfully the threshold $p_c$  by
the linear stability analysis of the stationary state.

For weak coupling, the threshold $p_c$ is likely
to remain near 1 as is seen for the example $a=(1+\sqrt{5}/2)$.
Thus, a tiny asynchronousnism may 
extinguish the collective motion.
Therefore, when GCM is used as model of real systems,
we keep in mind that 
even if the updating rule is almost synchronous,
the effect of asynchronous updating should not be ignored.

\section*{Acknowledgements}
This research was supported partly by Japan Society of Promotion of
Science under the contract number RFTF96I00102.

\end{document}